\documentclass{pos}

\usepackage{graphicx}

\newcommand{\qQED}{{\rm qQED}}

\newcommand{\ice}[1]{\relax}

\newcommand{\unl}{\underline}

\newcommand{\as}{a_s}

\newcommand{\beq}{\begin{equation}}

\newcommand{\ba}{\begin{array}}

\newcommand{\ea}{\end{array}}

\newcommand{\eeq}{\end{equation}}

\newcommand{\bea}{\begin{eqnarray}}

\newcommand{\eea}{\end{eqnarray}}

\title{Massless propagators: applications in QCD and QED}

\ShortTitle{Massless propagators:}

{ \vspace*{2mm}
\centerline{\normalsize\hfill  SFB/CPP-08-06}
\centerline{\normalsize\hfill  TTP08-03   }
\baselineskip 11pt
}

\author{{P.~A.~Baikov}\\
Institute of Nuclear Physics, Moscow State University,
Moscow~119899, Russia
\\
        E-mail: \email{baikov@theory.sinp.msu.ru}}
\author{\speaker{K.~G.~Chetyrkin}\\
      Institut f\"ur Theoretische Teilchenphysik,
  Universit\"at Karlsruhe, D-76128 Karlsruhe, Germany
 \\
        E-mail: \email{chet@particle.uni-karlsruhe.de}}
\author{J.~H.~K\"uhn\\
Institut f\"ur Theoretische Teilchenphysik,
  Universit\"at Karlsruhe, D-76128 Karlsruhe, Germany
\\
        E-mail: \email{jk@particle.uni-karlsruhe.de}}

\abstract{
We report on  two recent results based on the evaluation of  
five-loop massless propagators  in QCD and QED:
(i)  
corrections of order  $\alpha_s^4$ to the 
absorptive part of the polarization function  in QCD with 
$n_f=3$;
(ii)  the five-loop contribution to the $\beta$ function of  quenched  QED.

}

\FullConference{8th International Symposium on Radiative Corrections (RADCOR)\\
		 October 1-5 2007\\
		 Florence, Italy}

\begin{document}

\section{Introduction}
Propagators --- that is Feynman Integrals   depending on only one
external momenta --- appeared in QFT from its very beginning, for
instance, in the description of the vacuum polarization in QED and since then constitute
an important set of Feynman integrals to deal with.  

Massless propagators are  an  indispensable tool in Renormalization Group (RG) calculations
within the framework of Dimensional Regularization \cite{tHooft:1972fi} and the so-called
Minimal Subtractions  schemes \cite{tHooft:1973mm}.

Massless propagators  appear in many important physical applications.
The total cross-section of $e^+ e^- $ annihilation into hadrons, the
Higgs decay rate into hadrons, the semihadronic  decay rate of the
$\tau$ lepton,  the running of the fine structure coupling are all
computable in the high energy limit in terms of massless propagators.

The strong coupling constant $\alpha_s$ is one of the three
fundamental gauge couplings constants of the Standard Model  of
particle physics.  Its precise determination is one of the most
important aims of particle physics.
One of the most precise and theoretically safe determination 
of $\alpha_s$ is based on
measurements of the cross section for electron-positron annihilation
into hadrons. These have been performed in the low-energy region between 2~GeV
and 10~GeV and, in particular, at and around the $Z$ resonance at 
%PDG07: Mass m = 91.1876  €± 0.0021 GeV  
91.2~GeV. 
Conceptually closely related is the measurement of the semileptonic
decay rate of the $\tau$-lepton, leading to a determination of $\alpha_s$ at
a scale below 2 GeV.

%of %$m_\tau$
% PDG07: Mass m = 1776.99 +0:29 -  0:26 MeV
%$m_\tau=1.7776$~GeV and thus at relatively low energies

From the theoretical side, in the framework of perturbative QCD, these rates
and cross sections are evaluated as inclusive rates into massless quarks and gluons 
\cite{ChKK:Report:1996,Davier:2005xq}.
(Power suppressed  mass effects are  well under control for
$e^+e^-$-annihilation,  both at low energies and around the $Z$ resonance, 
and for $\tau$ decays 
\protect\cite{Chetyrkin:1996ii,Chetyrkin:1996cf,ChetKuhn90,ChK:mq4as2,ChKH:mq4as3,Baikov:2004ku}, 
and the same applies to mixed QCD and electroweak corrections \cite{Czarnecki:1996ei,Harlander:1997zb}).

The ratio 
$R(s)\equiv \sigma(e^+e^-\to {\rm hadrons}) / \sigma(e^+e^-\to\mu^+\mu^-)$ 
is 
%conveniently 
expressed through the absorptive part of the 
correlator 
\begin{equation}
\label{Pi}
\Pi_{\mu\nu}(q)  =
 i \int {\rm d} x e^{iqx}
\langle 0|T[ \;
\;j_{\mu}^{\rm em}(x)j_{\nu}^{\rm em}(0)\;]|0 \rangle
=
\displaystyle
(-g_{\mu\nu}q^2  + q_{\mu}q_{\nu} )\Pi(-q^2)
{}\, ,
\end{equation}
with the hadronic EM current
$
j^{\rm em}_{\mu}=\sum_{{f}} Q_{{f}}
\overline{\psi}_{{f}} \gamma_{\mu} \psi_f
$, and $Q_f$ being the EM charge of the quark $f$.
The optical theorem
relates  the inclusive cross-section
and thus the function $R(s)$
to the discontinuity of $\Pi$
in the complex plane
\begin{equation} \label{d3}
R(s) =  \displaystyle
 12 \,\pi \,{\rm Im}\, \Pi( - s -i\delta)
\label{discontinuity}
{}\, .
\end{equation}
The renormalization mode of the polarization operator
$\Pi(Q^2)$  reads  (see, e.g. ~\cite{ChKK:Report:1996})
\beq
1 + e^2\, \Pi(Q^2/\mu^2,\alpha_s) = Z^{\rm ph}_3  + e^2 \, \Pi_0(Q^2,\alpha_s^0),
{},
\label{renorm:mod}
\eeq
where $e = \sqrt{\alpha\, 4\pi}$  and $\alpha$ is the  (renormalazied) fine structure constant.
Eq. (\ref{renorm:mod}) can be  naturally deduced from the  connection
between  $\Pi(Q^2)$
and  the  photon propagator $D_{\mu\nu}(q)$
\[
D_{\mu\nu}(q) = g_{\mu\nu}\frac{i}{q^2}\frac{1}{1+e^2\,\Pi(q^2=-Q^2)}
{}.
\]
It is also convenient to introduce  the Adler function as
\beq
\nonumber
{D}(Q^2) =  -12\, \pi^2
Q^2\, \frac{\mathrm{d}}{\mathrm{d} Q^2} \Pi(Q^2)
=
\int_0^\infty \frac{ Q^2\ R(s) d s }{(s+Q^2)^2}
{}.
\eeq
We define  the perturbative  expansions 
\beq
{D}(Q^2) = \sum_{i=0}^{\infty} \  {d}_i a_s^i(Q^2),  \ \
{R}(s) =  \sum_{i=0}^{\infty} \  {r}_i a_s^i(s)
{},  
%\nonumber
\eeq
where $a_s \equiv \alpha_s/\pi$ and  we have set the normalization scale $\mu^2=Q^2$ or o $\mu^2= s$
for the Euclidian and Minkowskian functions respectively.
The results
for generic values of $\mu$ can be easily recovered with  standard
RG techniques.

For the vector correlator the terms of order $a_s^2$ and $a_s^3$ have been
evaluated nearly thirty and about fifteen years ago 
\cite{Chetyrkin:1979bj,Gorishnii:1991vf}, respectively. 
The $a_s^4$ corrections are conveniently classified according to their 
power of $n_f$, with $n_f$ denoting the number of light quarks.  
The $a_s^4 n_f^3$ term is part of the ``renormalon chain'', the evaluation of
the next term, of order $a_s^4 n_f^2$, was a test case for the techniques used
extensively in this paper and, furthermore, led to useful
insights into the structure of the perturbative series already 
\cite{ChBK:vv:as4nf2}. The rest: two remaining most difficult terms of orders  $a_s^4 n_f^2$ 
and  $a_s^4 n_f^0$  of equivalent complexity\footnote{
In the sense that they both contain all possible topologies.}
are not yet known. The results of their (partial) evaluation 
will be presented in the talk.

\section{Calculation of $d_4$ at $n_f=3$} 

The complete five-loop calculation requires the evaluation of 2671 Abelian
quenched  plus about seventeen thousand non-abelian and
non-quenched diagrams (see Fig 1). 
\vspace{.1cm}

\begin{figure}[h]
\begin{center}
\includegraphics[width=3cm]{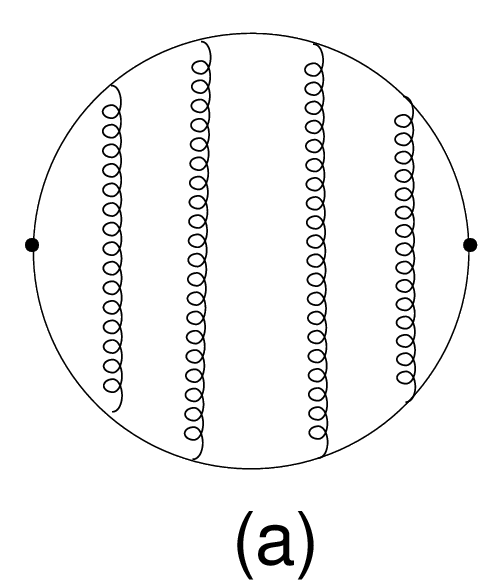}
\hfill
\includegraphics[width=3cm]{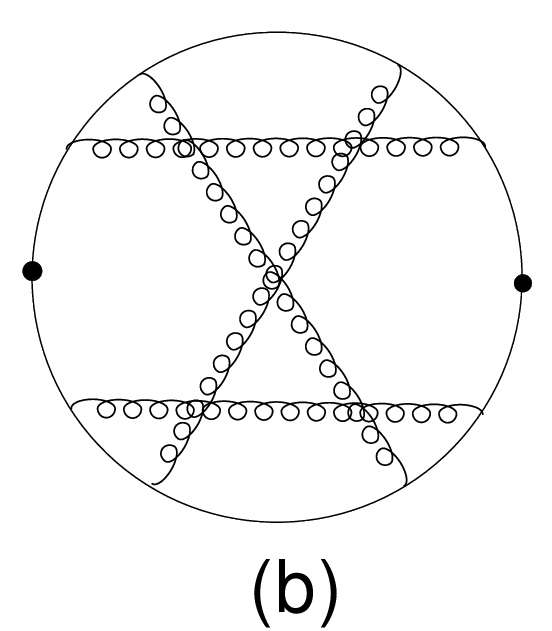}
\hfill
\includegraphics[width=3.7cm]{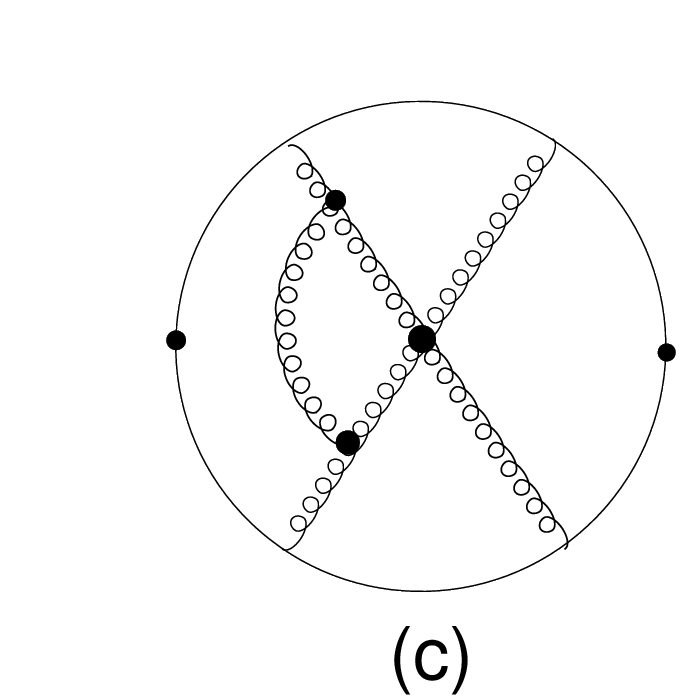}
\hfill
\includegraphics[width=3cm]{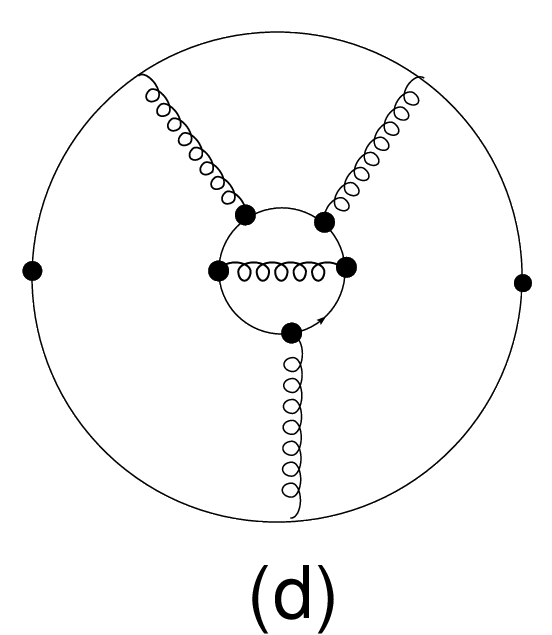}
\end{center}
\caption{(a) (a),(b)  Examples of five -loop  quenched  Abelian diagrams of order $a_s^4 n_f^0$ .
(c) A  five -loop  non-abelian diagram.
(d) A  five -loop  non-quenched  abelian diagram of order $a_s^4 n_f$
}
\end{figure}
Using ``infrared rearrangement''
\cite{Vladimirov:1980zm}, the
$R^*$ operation \cite{ChS:R*} and the prescriptions formulated in
\cite{gvvq} to algorithmically resolve the necessary combinatorics, it
is possible to express the absorptive part of the five-loop diagrams
in terms of four-loop massless propagator integrals
(``p-integrals''). Using then a representation for Feynman integrals
proposed in \cite{Baikov:tadpoles:96,Baikov:explit_solutions:97},  these
p-integrals can be reduced to a sum of 28 master integrals, with
coefficients which are rational functions of the space-time dimension
$D$. These coefficients were, in a first step, evaluated in the
large-$D$ limit, and, after evaluating sufficiently many terms of the
$1/D$ expansion, fully reconstructed \cite{Baikov:2005nv}.  
This direct and largely automatic procedure required enormous computing resources and
was performed on a 32+8-node SGI parallel computer with SMP architecture
and on the HP XC4000 supercomputer of the federal state   Baden-W\"urttemberg 
%(claster of Dual Core AMD Opteron 2.6 GH),
using a parallel version \cite{Tentyukov:2004hz} of FORM 
\cite{Vermaseren:2000nd}.

The construction of the large-$D$ limit requires in general huge 
storage resources, which naturally constrains the structure of the 
input p-integrals: they should better not contain any  extra parameters like
color coefficients, $n_f$--the number of light quark  flavours contributing to  internal fermion
loops--, and so on. As a result we are forced to use a ``slice'' approach:
that is to set all color   coefficients  to their  numerical  values and 
to fix $n_f$ to some integer. Thus, in order to compute the remaining
two not-yet-known $\alpha_s^4$ contributions to $R(s)$   one should  compute
two slices: $n_f=0$ and $n_f=n_0$. Here $n_0$ stands for any non-zero 
integer which could be chosen at will. Since the  evaluation of every single slice 
is a problem  by itself, we decided to start from $n_0=3$ as  the result 
has important physical applications for  the analysis of the QCD corrections
to  hadronic  $\tau$-decays (see, e.g. \cite{Davier:2005xq}). 

Our result reads (we suppress the trivial factor $3\,\sum_f Q^2_f$
throughout\footnote{
In this paper we present the results for the so-called ``non-singlet''
diagrams, where one and the same closed  quark line is connected to the external
currents. These are sufficient for a complete description of
$\tau$-decays. For $e^+e^-$ annihilation they correspond to the dominant terms
proportional $\sum_f Q_f^2$. The singlet contributions proportional
$(\sum_f Q_f)^2$  arise for the first time in ${\cal O}(\alpha_s^3)$. They
are known to be small, and will be evaluated at a later
point.
})
\bea
 d_4(n_F =3) &=&
 \frac{78631453}{20736} - \frac{1704247}{432} \,\zeta_3
+ \frac{4185}{8}\,\zeta_3^2   + \frac{34165}{96}\,\zeta_5 -
      \frac{1995}{16}\,\zeta_7
\\
&\approx&  49.0757
{}.
\eea
The corresponding expression for $R(s)$ is:
\beq
1 + \as + 1.6398\,\as^2 + 6.3710\,\as^3 \,- 106.8798
\,\as^4
\label{R_nf3_as4}
{}.
\eeq

\ice{
It is  instructive to explicitly display  the genuine five-loop contributions to  $d_3$ and $d_4$
(underlined in (\ref{r_3_separated},\ref{r_4_separated}) below) and  the  ``kinematical'' terms
originating from the analytic continuation:
\bea
r_3 &=& \unl{18.2} - 24.9  + (\unl{-4.22} + 3.02)\, n_f
+ (\unl{-0.086} + 0.091)\, n_f^2,
\label{r_3_separated}
\\
r_4(n_f=3)  &=& \unl{49.08}  - 155.956
\label{r_4_separated}
{}.
\eea
}
Since it will presumably take a long time until the next term of the
perturbative series will be evaluated,
%exactly,
it is of interest to investigate
the predictive power of various optimization schemes empirically. Using the
principles of ``Fastest Apparent Convergence'' (FAC)  \cite{Grunberg:1984fw}
or of ``Minimal
Sensitivity'' (PMS) \cite{Stevenson:1981vj}, which happen to coincide in this order, 
one gets \cite{Kataev:1995vh,ChBK:tau:as4nf2}
$$
d_4^{\rm pred}(n_f=3) = 27 \pm 16
{},
$$
with the central value of 
differing  significantly from the exact result
\beq
d_4^{\rm exact}(n_f=3) = 49.08
{}.
\eeq
However, within the error estimates \cite{ChBK:tau:as4nf2},  predicted and exact values are
in agreement.
%while  the deviations are basically in agreement to the  estimated
%in  \cite{ChBK:tau:as4nf2}  accuracy  of  the FAC/PMS predictions.
%
The picture changes, once these estimates are used to predict the
coefficient $r_4$.
\ice{
 Although sizable cancellations between ``dynamical'' and
``kinematical'' terms are observed for the individual $n_f$ coefficients in
(\ref{r_4_separated})
}
The prediction  for the final result
is  significantly closer (in a  relative sense) to the result of the exact calculation:
\beq
r_4^{\rm pred}(n_f=3) =
-129 \pm 16
{},  \  \   
\nonumber
r_4^{\rm exact}(n_f=3) = - 106.88
{}.
\eeq
This is in striking contrast to the case  of the scalar correlator,  where the
predictions for the dynamical terms work well, but, as a consequence of the
strong cancellations between dynamical and kinematical terms fail completely
in the Minkowskian region \cite{Baikov:2005rw}.

\section{Colour structure of $d_4$}

In general, it would be of great  interest  to know the $\alpha_s^4$ contribution 
for a generic color group. The structure of the result for, say, Adler function, may be
predicted from \cite{Vermaseren:1997fq}: 
\beq
d_4 = \sum_{i=1,12}\,  d_{4,i}\, c_{4,i}
{},
\eeq
with 
the coefficients $d_{4,i}$ being polynomials in ${\rm ln }(\frac{\mu^2}{Q^2})$ and the following 
colour factors 
read
\beq
\begin{array}{c} 
 c_4 =  
\{
d_R\, C_F^4\,,\, d_R\, C_F^3 C_A\,,\, d_R\, C_F^2 C_A^2 \,,\,d_R\, C_F C_A^3 \,,\, d_R\,C_F^3 T_F n_f\,,\,d_R\, C_F^2 C_A T_F n_f\,,
%\right.
\\
%\left.
d_R\,C_F C_A^2 T_F n_f\,,\,  d_R\,C_F^2 T_F^2 n_f^2\,,\,  d_R\,C_F C_A T_F^2 n_f^2\,,\, d_R\,C_F T_F^3 n_f^3\,,\, 
{d_F^{a b c d} d_A^{a b c d}}\,
 , \,
 n_f {d_F^{a b c d}d_F^{a b c d}}
\}
{}.
\end{array}
\eeq
Here 
 $[T^a T^a]_{ij} = C_F \delta_{ij}$ and
 $f^{a c d} f^{b c d} = C_A \delta^{ab}$ are the quadratic Casimir
operators of the
fundamental and the adjoint representation of the Lie algebra and
tr$(T^a T^b) = T_F \delta^{a b}$ is the trace normalization of the
fundamental representation. $d_R$ is the dimension of the fermion
representation (i.e. the number of quark colours) and $n_f$ is the number of
quark flavors. The exact definitions for the higher order 
group invariants ${d_F^{a b c d} d_A^{a b c d}}$ and ${d_F^{a b c d}d_F^{a b c d}}$
are given in \cite{Vermaseren:1997fq}.
For  QCD with \ice{colour gauge group} SU(3):
\[
C_F =4/3\,,\, C_A=3\,,\, T_F=1/2\,,\, d_R = 3\,,\, 
{d_F^{a b c d} d_A^{a b c d}} = \frac{15}{2}\,,\, {d_F^{a b c d}d_F^{a b c d}} = \frac{5}{12}
{}.
\]
We observe that to  get  the full colour structure of $d_4$ we should compute as many as 
eight extra slices (as four slices are already available: $d_{4,10}$ \cite{Beneke:1992ch}, $  d_{4,9}$ and
$  d_{4,8}$ \cite{ChBK:vv:as4nf2}, as well as    $SU(3)$ with $n_f=3$). Thus, complete  evaluation of $d_4$ for generic
colour group  is  a matter of future. 

However, not all slices are equally difficult.
For  the case of the abelian U(1) group (QED!) the colour coefficients have especially simple form:
 \[
C_F =1\,,\, C_A=0\,,\, T_F=1\,,\, d_R = 1\,,\, 
{d_F^{a b c d} d_A^{a b c d}} = 0\,,\, {d_F^{a b c d} d_F^{a b c d}} = 1
{}.
\]
In addition the corresponding  diagrams are somewhat simpler due to the absence  of the
three- and four-gluon couplings as well as any  vertexes with ghost lines.  An even simpler slice
corresponds to case of  quenched  QED (qQED):
 \[
C_F =1\,,\, C_A=0\,,\, T_F=1\,,\, d_R = 1\,,\,n_f= 0\,, \, 
{d_F^{a b c d} d_A^{a b c d}} = 0\,,\, {d_F^{a b c d} d_F^{a b c d}} = 0
{}.
\]
Thus, in general the   quenched  QED contribution to   $d_n$ is equal  to the coefficient
in front of the  $d_R\, (C_F)^n$ structure in the  colour decomposition of $d_n$. 

Up to order $\alpha^3$ the Adler function for qQED  reads \cite{Gorishnii:1991vf} 
\beq
D^{\qQED}(\alpha,Q) = 1 + 3\,A -\frac{3}{2}\,A^2 -\frac{69}{2}\, A^3
{},
\label{DqQED}
\eeq 
where $A = A(\mu) = \frac{\alpha(\mu)}{4\,\pi}$. Note that
$D^{\qQED}(\alpha,\mu/Q)$ does not depend on its second argument  at all, which means that it
is an essentially finite, scheme-independent  quantity. It is instructive to compare
eq. (\ref{DqQED} ) to the $\beta$-function of qQED, also known to the
same order from \cite{Rosner66,Rosner67,Gorishnii:1990kd}:
\beq
\beta^{\qQED} =     \frac{4}{3}\, A\, \left\{
1 + 3\,A -\frac{3}{2}\,A^2 -\frac{69}{2}\, A^3
\right\}
\label{betaqQED}
{}.
\eeq 
One observes the validity of the relation
\beq
D^{\qQED}(\alpha,Q) = \frac{4}{3}\, A\, {\bf \beta}^{\qQED}
{}
\eeq 
which can be easily  proven  in all orders  of perturbation theory starting from (\ref{renorm:mod}). 
Thus, the  evaluation of the five-loop contribution to ${\bf \beta}^{\qQED}$ provides with us another slice, namely
with $d_{4,1}$. 

On the other hand,  such a calculation is of importance  by  itself. 
Indeed, the $\beta$- function of  qQED possesses a number of remarkable and intriguing features.
\begin{itemize}

\item It  is gauge and scheme independent in all orders.

\item Its coefficients are simple rational numbers at three and four loops\footnote{
The rationality at one and two loops  takes  place for any $\beta$-function irrespectively
of the (massless) theory.}. 

\item If $\beta^{\qQED}(\alpha_0) = 0$
then $\alpha = \alpha_0$ leads to a self-consistent {\bf finite}
solution  of (massless) QED \cite{Johnson:1967pk,Johnson:1973pd}
%en{/K.Johnson and M. Baker, (1973)/}

\end{itemize}

A detailed knot-theoretic explanation of the
rationality property at three loop level was given in 
\cite{Broadhurst:1995dq,Broadhurst:1999zi}. At four loop the
problem  was thoroughly investigated with the  help of a dimensionally 
background-field method  in \cite{Broadhurst:1999zi,Broadhurst:1999xk}. 
Some hope that the  rationality property is not  accidental  but  also holds
in  higher orders has been expressed in   \cite{Broadhurst:1999xk}. 
Unfortunately, no clear unambiguous prediction  as for the  structure
of   higher (five loops and beyond)  orders has ever been made.

We have computed $d_{4,1}$. Our five loop result for the  $\beta$ function reads:  
\beq
{\bf \beta}^{\qQED} = \frac{4}{3}\, A
 + 4\,A^2 -{2}\,A^3 -46 \, A^4 + 
\left(\frac{4157}{6}  \ \  + \ \ 128\,  \zeta_3\right)\, A^5
\label{betaqQED_5l}
{}.
\eeq 
Thus our  calculation of $d_{4,1}$ has solved the problem: the rationality ceases to exist
starting from  fifth loop.

{ \em Acknowledgments.}
We thank  David Broadhurst and Dirk Kreimer 
for useful discussions.
This work was supported by
the Deutsche Forschungsgemeinschaft in the
Sonderforschungsbereich/Transregio
SFB/TR-9 ``Computational Particle Physics'',  by INTAS (grant
03-51-4007) and by RFBR (grant 05-02-17645).
The computer  calculations were partially  performed on  the  HP XC4000  super
computer of the federal state   Baden-W\"urttemberg at the High Performance Computing Center Stuttgart 
(HLRS) under the grant ``ParFORM''.

{\em Note added.} Recently we have finished the calculation of $d_4$ for a generic value of $n_f$.
The reader is referred to  \cite{Baikov:2008jh} for  details and for the discussion of phenomenological
applications of the result.

\end{document}